\documentclass{natureWMF}

\usepackage{graphicx}

\newcommand\beq{\begin{equation}}
\newcommand\eeq{\end{equation}}

\newcommand{\Ham}{{\mathcal H}}

\newcommand{\tot}{\mathrm{tot}}

\long\def\symbolfootnote[#1]#2{\begingroup\def\thefootnote{\fnsymbol{footnote}}\footnote[#1]{#2}\endgroup}
\makeatletter
\title{Hot Jupiters from Secular Planet--Planet Interactions}
\author{\bf Smadar Naoz$^{1}$, Will M. Farr$^{1}$, Yoram Lithwick$^{1}$, Frederic A.
Rasio$^{1}$, Jean Teyssandier$^{1}$}
\begin{document}
\maketitle
\begin{affiliations}
\item  Center for Interdisciplinary Exploration and Research in Astrophysics (CIERA),  
Northwestern University, Evanston, IL 60208,
USA

\end{affiliations}

\begin{abstract}
  About 25 per cent of ‘hot Jupiters’ (extrasolar Jovian-mass planets with close-in orbits) are actually orbiting counter to the spin direction of the star\cite{Tri+10}. Perturbations from a distant binary
  star companion\cite{Dan,Wu+07} can produce high inclinations, but cannot explain orbits that are retrograde with respect to the total angular momentum of the system. Such orbits in a stellar context can be produced through secular (that is, long term) perturbations in hierarchical triple-star systems.   Here we report a similar application to
  planetary bodies, including both the key octupole-order effects and
  tidal friction, and find that it can produce hot Jupiters in orbits
  that are retrograde with respect to the total angular momentum.
  With  distant stellar mass perturbers such an outcome is not
  possible\cite{Dan,Wu+07}.  With planetary perturbers the inner
  orbit's angular momentum component parallel to the total angular
  momentum need not be constant\cite{Ford+00}.  In fact, as we show
  here, it can even change sign, leading to a retrograde orbit.  A
  brief excursion to very high eccentricity during the chaotic
  evolution of the inner orbit can then lead to rapid capture, forming
  a retrograde hot Jupiter.
\end{abstract}

Despite many
attempts\cite{Dan,Wu+07,Sourav+08,Lai+10,Nag+08,Schlaufman,Takeda,Winn+10b,WL},
there is no model that can account for all the properties of the known
hot Jupiter (HJ) systems.  One model suggests that HJs formed far away
from the star and slowly spiraled in, losing angular momentum and
orbital energy to the protoplanetary disk\cite{Lin+86,Mass+03}.  This
``migration'' process should produce planets with low orbital
inclinations and eccentricities. However, many HJs are observed to be
on orbits with high eccentricities, and misaligned with the spin direction
of the star (as measured through the Rossiter--McLaughlin
effect\cite{GW07}) and some of these ($8$ out of $32$) even appear to
be orbiting counter to the spin of the star.  In a second model,
secular perturbations from a distant binary star companion can produce
increases in the eccentricity and inclination of a planetary
orbit\cite{Hol+97}.  During the evolution to high eccentricity, tidal
dissipation near pericenter can force the planet's orbit to decay,
potentially forming a misaligned HJ\cite{Dan,Wu+07}. Recently, secular
chaos involving several planets has also been proposed as a way to
form HJs on eccentric and misaligned orbits\cite{WL}.  A different
class of models to produce a tilted orbit is via planet--planet
scattering\cite{Sourav+08}, possibly combined with other perturbers
and tidal friction\cite{Nag+08}. In such models the initial
configuration is a densly-packed system of planets and the final
tilted orbit is a result of dynamical scattering among the planets, in
contrast to the secular interactions we study here.

In our  general treatment of secular interactions  between two orbiting bodies
 we allow for the magnitude and orientation of
{\it both} orbital angular momenta to change (see Figure~1).  The outer
body (here either a planet or a brown-dwarf)
gravitationally perturbs the inner planet on time scales long compared
to the orbital period (i.e., we consider the secular evolution of the
system). We define the orientation of the inner orbit with respect to the
invariable plane of the system (perpendicular to the {\em
  total\/} angular momentum): a
prograde (retrograde) orbit has $i_1<90^\circ$ ($i_1>90^\circ$), where
$i_1$ is the inclination of the inner orbit with respect to the
  total angular momentum vector. Note that the word ``retrograde" is also used in the literature 
  to indicate orbital motion  counter to the stellar spin. The directly observed parameter is actually
  the {\it projected} angle between the spin axis of the star and
  the orbital angular momentum of a HJ.  
  Our proposed mechanism can produce HJs that are ``retrograde"
 {\it both} with respect to the stellar spin and with respect to the
  total angular momentum.  By contrast, a distant stellar companion
  can only succeed in the former. See the online Supplementary Information 
  for more details; henceforth we will use the term ``retrograde" only to indicate an 
  orbit with $i_1>90^\circ$ as define above.
    
We assume a hierarchical
configuration, with the outer perturber on a much wider orbit
than the inner one. In the secular approximation the orbits may change
shape and orientation but the semi-major axes are strictly
conserved in the absence of tidal
dissipation\cite{Ford+00,1998KEM}. In particular, the
Kozai-Lidov mechanism\cite{Kozai,Lidov,Mazeh+79}
produces large-amplitude oscillations of the eccentricity and
inclination when the initial relative inclination between the inner
and outer orbits is sufficiently large ( $40^\circ < i <
140^{\circ}$).

We have derived the secular evolution equations to octupole order
using Hamiltonian perturbation
theory\cite{Harr69,Mazeh+99,Ford+00}.  In contrast
to previous derivations of ``Kozai-type'' evolution, our treatment
allows for changes in the $z$-components of the orbital angular
momenta (i.e., the components along the total angular momentum)
$L_{{\rm z},1}$ and $L_{{\rm z},2}$ (see Supplementary Information). The
octupole-order equations allow us to calculate the
evolution of systems with more closely coupled orbits and with
planetary-mass perturbers. The octupole-level terms can give rise to
fluctuations in the eccentricity maxima to arbitrarily high
values\cite{Ford+00,Mazeh+99}, in contrast to the regular
evolution in the quadrupole
potential\cite{Dan,Wu+07,Mazeh+79}, where the
amplitude of eccentricity oscillations is constant.  

Many previous
studies of secular perturbations in hierarchical triples considered a
stellar-mass perturber, for which $L_{\rm z,1}$ is very nearly
constant\cite{Dan,Wu+07,Mazeh+79}.  Moreover, the
assumption that $L_{\rm z,1}$ is constant has been built into previous
derivations\cite{1998EKH,Mik+98}.
However, this assumption is only valid as long as $L_{2} \gg L_{1}$,
which is not the case in comparable-mass systems (e.g., with two
planets). Unfortunately, an immediate consequence of this assumption
is that an orbit that is prograde relative to the total angular momentum
always remains prograde.
Figure \ref{fig:point-mass-dynamics} shows the evolution of a
representative system (here without tidal effects for simplicity): the
inner planet oscillates between prograde and retrograde orbits
(with respect to the total angular momentum) as
angular momentum flows back and forth between the two orbits.

Previous calculations of planet migration through ``Kozai cycles with
tidal friction''\cite{Dan,Wu+07,1998KEM,Mazeh+79}
produced a slow, gradual spiral-in of the inner planet. Instead, our
 treatment shows that the eccentricity can occasionally
reach a much higher value than in the regular ``Kozai cycles''
calculated to quadrupole order. Thus, the pericenter distance will
occasionally shrink on a short time scale (compared to the Kozai
period), and the planet can then suddenly be tidally captured
by the star.  We propose to call this ``Kozai capture.'' 

Kozai capture provides a new way to form HJs. If the capture happens after the inner
orbit has flipped the HJ will appear in a retrograde orbit. This is
illustrated in Figure~2.  During the evolution of the system the inner
orbit shrinks in steps (Fig.~2c) whenever the dissipation becomes
significant, i.e., near unusually high eccentricity maxima.  The inner
orbit can then eventually become tidally circularized.  This happens
near the end of the evolution, on a very short time scale (see Fig.~2,
right panels). In this final step, the inner orbit completely and
quickly decouples from the outer perturber, and the orbital angular
momenta then become constant.  Therefore, the final semi-major axis for the HJ is
$\approx 2 r_p $, where $r_p$ is the pericenter distance at the
beginning of the capture phase\cite{Ford+06}.

The same type of evolution shown in Figure~2 is seen with a broad
range of initial conditions.  There are two main routes to forming a
HJ through the dynamical evolution of the systems we consider here. In
the first, tidal friction slowly damps the growing eccentricity of the
inner planet, resulting in circularized, prograde HJs. In the second,
a sudden high-eccentricity spike in the orbital evolution of the inner
planet is accompanied by a flip of its orbit. The planet is then
quickly circularized into a retrograde short-period orbit. We can
estimate the relative frequencies of these two types of outcomes using
Monte Carlo simulations. Given the vast parameter space for initial
conditions, a complete study of the statistics is beyond the scope of
this Letter (but see Naoz et al., in preparation).  However, we can
provide a representative example: consider systems where the inner
planet was formed {\it in situ\/} at $a_1=5\,$AU with zero obliquity
(orbit in the stellar equatorial plane) and with some small
eccentricity $e_1=0.01$, while the outer planet has $a_2= 51
\,$AU. The masses are $m_1=1 \,M_{\rm J}$ and $m_2=3 \,M_{\rm J}$.  We
draw the eccentricity of the outer orbit from a uniform distribution
and the mutual inclination from a distribution uniform in $\cos i$
between $0$ and $1$ (i.e., isotropic among prograde orbits). For this
case we find that, among all HJs that are formed, about $7\%$ are in
truly retrograde motion (i.e., with respect to the total angular
momentum) and about $50\%$ are orbiting counter to the stellar spin
direction.  Note that the latter fraction is significantly larger than what
previous studies have obtained with stellar-mass perturbers (at most
$\sim 10\%$ \cite{Dan,Wu+07}).  The high observed incidence of planets
orbiting counter to the stellar spin direction\cite{Tri+10} may suggest
that planet--planet secular interactions are an important part of
their dynamical history.

 Our mechanism requires that two coupled orbits
 start with a relatively high mutual inclination ($i > 50^\circ$).
The particular configuration in Figure~2 has a very wide outer orbit
similar to those of directly imaged planets such as
Fomalhaut~b\cite{Kal+08} and HR~8799b\cite{Mar+08}. In this case the
inner Jupiter could have formed in its original location in accordance
with the standard core accretion model\cite{Pol+96} on a nearly circular orbit.
An alternative
path to such a configuration involves strong planet--planet scattering
in a closely packed initial system of several giant
planets\cite{Nag+08}.  Independent of any particular planet formation
mechanism, we predict that systems with misaligned HJs should also
contain a much more distant massive planet or brown dwarf on an inclined orbit.

\newpage

\begin{figure*}[hptb]
\includegraphics[width=\textwidth]{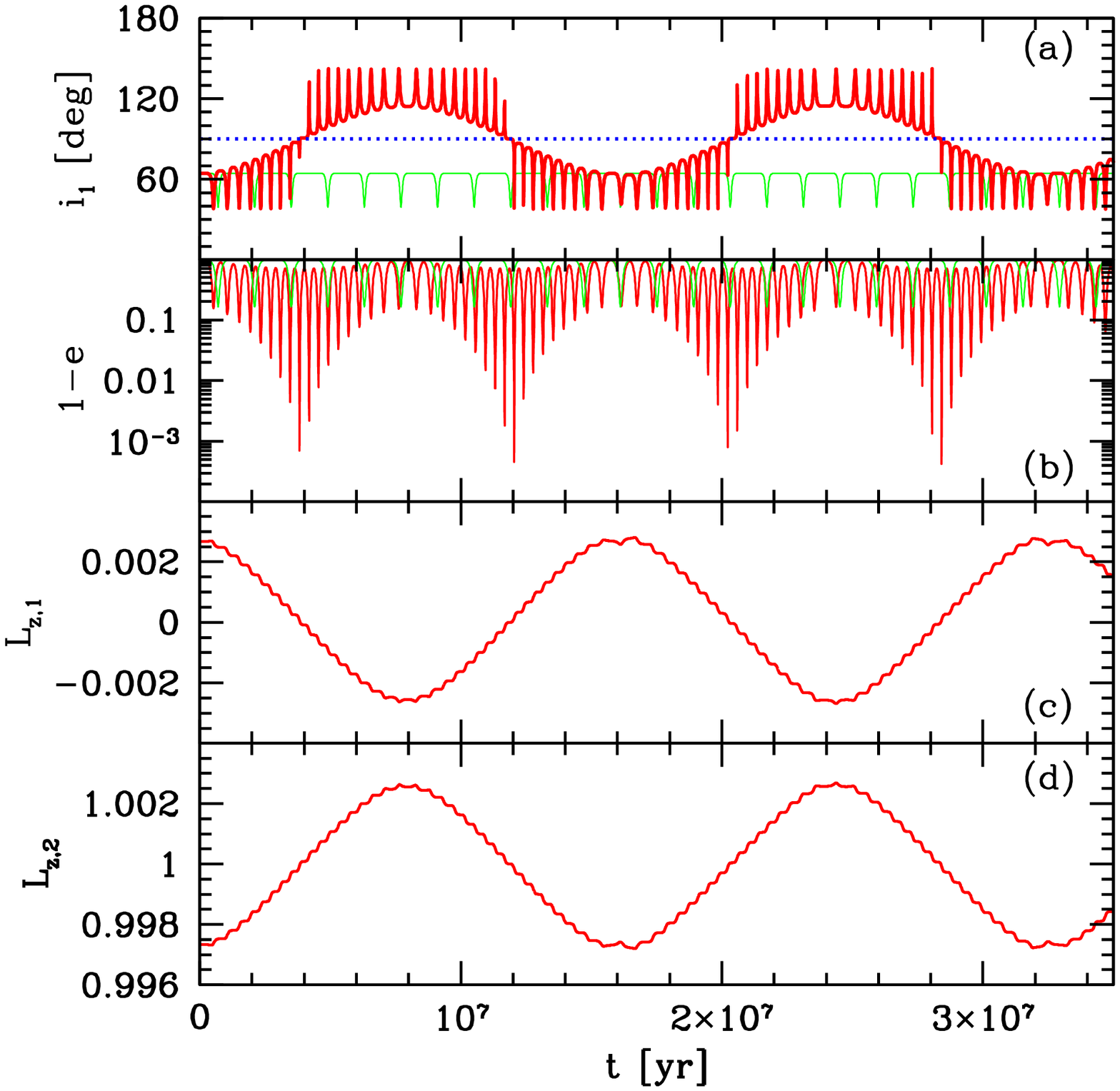}
\end{figure*}
\begin{figure*}[hptb]
  \caption{\label{fig:point-mass-dynamics} \sffamily \small{ {\bf Dynamical
      evolution of a representative planet and brown dwarf system.}
       Here we ignore tidal dissipation, but we do include the
lowest-order
post-Newtonian precession rate for the inner orbit. Here the star has mass $1\,M_\odot$, the 
      planet $1\,M_{\rm J}$ and the outer brown dwarf $40\,M_{\rm J}$.
      The inner orbit has $a_1 = 6\,$AU and the outer orbit has
       $a_2 = 100\,$AU. The initial eccentricities are $e_1 =0.001$
      and $e_2 =0.6$ and the initial relative inclination
      $i=65^\circ$.  We show from top to bottom: (a) the inner orbit's
      inclination ($i_1$); (b) the eccentricity of the inner orbit (as
      $1 - e_1$); (c) and (d) the $z$-component of the inner- and
      outer-orbit's angular momentum, normalized to the total angular
      momentum (where the $z$-axis is defined to be along the total
      angular momentum).  The thin horizontal line in (a) marks the
      $90^\circ$ boundary, separating prograde and retrograde
      orbits. The initial mutual inclination of $65^\circ$ corresponds
      to an inner and outer inclination with respect to the total angular momentum
       (parallel to $z$) of $64.7^\circ$ and $0.3^\circ$,
      respectively. During the evolution, the eccentricity and inclination
      of the inner orbit oscillate, but, in contrast to what would be
      predicted from evolution equations truncated to quadrupole order
      [shown by the thin curves in panels (a) and (b)], the
      eccentricity of the inner orbit can occasionally reach extremely
      high values and its inclination can become higher then
      $90^\circ$.  The outer orbit's inclination always remains near
      its initial value.  We note that more compact systems usually do
      not exhibit the same kind of regular oscillations between
      retrograde and prograde orbits illustrated here, as chaotic
      effects become more important and are revealed at octupole order
      (see Fig.~2). We find that $\sim 50\%$ of the time the inner orbit is retrograde.  
      }}
\end{figure*}

\begin{figure*}[hptb]
\includegraphics[width=\textwidth]{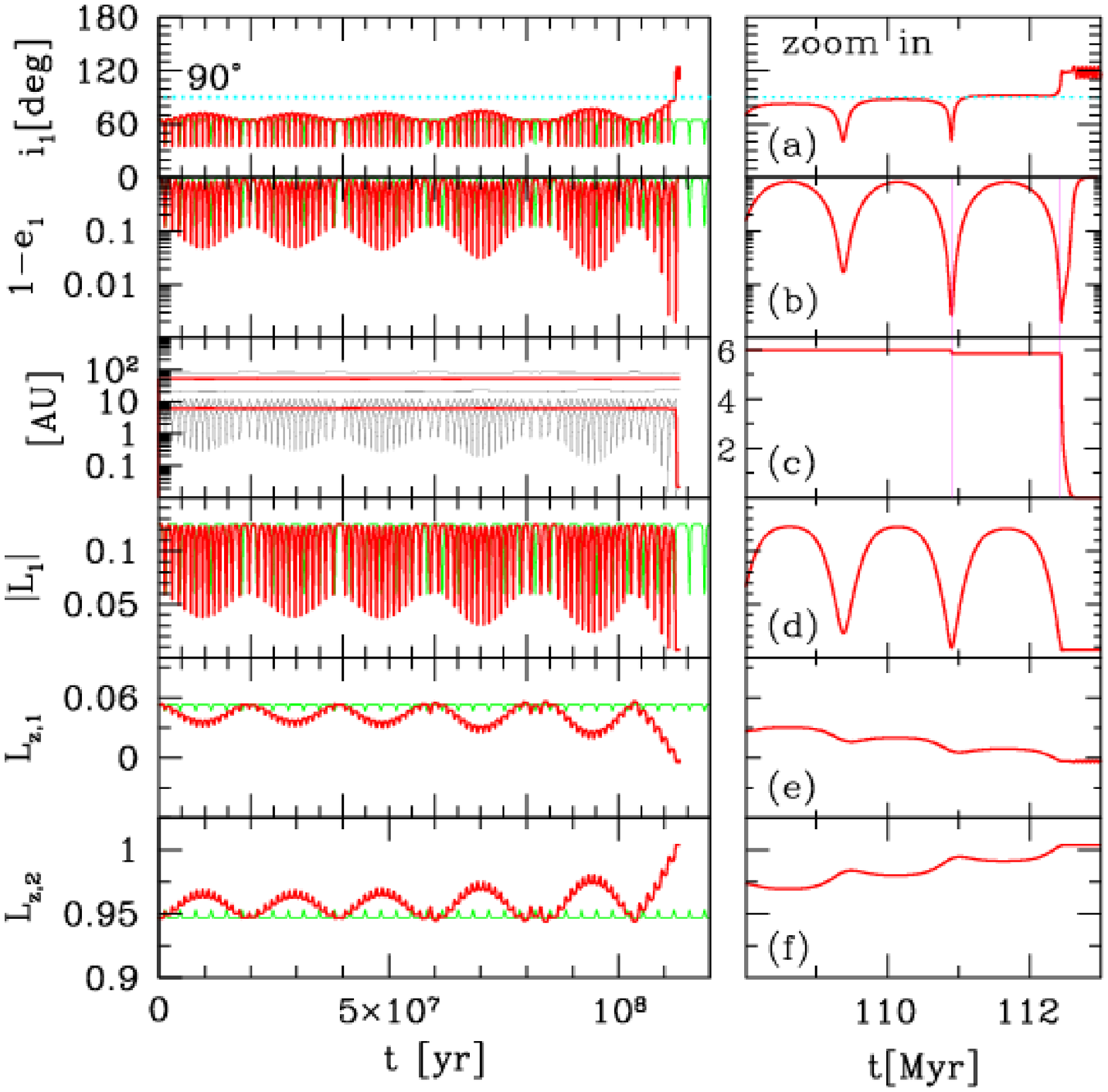}
\end{figure*}
\begin{figure*}[hptb]
  \caption{\label{fig:HJ-capture}\sffamily \small{ {\bf Dynamical evolution of a
      representative two-planet system with tidal dissipation
      included}.  The inner planet becomes retrograde at 112 Myr, and
      remains retrograde after circularizing into a HJ.  Here
      the star has mass $1\,M_\odot$, the inner planet $1\,M_{\rm J}$
      and the outer planet $3\,M_{\rm J}$.  The inner orbit has 
      $a_1 = 6\,$AU and the outer orbit has  $a_2 = 61\,$AU.  The
      initial eccentricities are $e_1 =0.01$ and $e_2 =0.6$, the
      initial relative inclination $i=71.5^\circ$, and argument of periapsis is $45^\circ$.  We show: (a) the
      inner orbit's inclination ($i_1$); (b) the eccentricity of the
      inner orbit (as $1 - e_1$); (c) the semi-major axis, peri-, and apo-center
      distances for the inner orbit and the peri- and apo-center
      distances for the outer orbit; (d) the magnitude of the angular
      momentum of the inner orbit; and, in (e) and (f) the
      $z$-components of the inner and outer orbit's angular momenta,
      normalized to the total angular momentum. The initial mutual
      inclination of $71.5^\circ$ corresponds to inner- and outer-orbit
      inclinations of $64.7^\circ$ and $6.8^\circ$,
      respectively. During each excursion to very high eccentricity
      for the inner orbit [marked with vertical lines in panels (b)
      and (c)], tidal dissipation becomes significant. Eventually the
      inner planet is tidally captured by the star and its orbit
      becomes decoupled from the outer body. After this point the
      orbital angular momenta remain nearly constant.  The final semi-major axis
      for the inner planet is  $0.022\,$AU, typical for a 
      HJ. The thin curves in panels (a),(b),(d),(e) and (f) show the evolution in the
      quadrupole approximation (but including tidal friction),
      demonstrating that the octupole-order effects lead to a
      qualitatively different behavior. For the tidal evolution in
      this example we assume tidal quality factors $Q_\star =5.5
      \times 10^6$ for the star and $Q_{\rm J}=5.8\times 10^6$ for the
      HJ (see Supplementary Information). 
      We monitor the pericenter distance of the inner planet to ensure
      that it always remains outside the Roche limit\cite{Soko}.  Here, as in Figure 1, we also include the 
      lowest-order post-Newtonian precession rate for the inner orbit. 
      }}
\end{figure*}

\newpage


\begin{addendum}
 \item We would like to thank Dan Fabrycky and Hagai Perets for
   discussions. SN acknowledges support from a Gruber Foundation
   Fellowship and from the National Post Doctoral Award Program for
   Advancing Women in Science (Weizmann Institute of
   Science). Simulations for this project were performed on the HPC
   cluster {\it{fugu}} funded by an NSF MRI award.
 \item[Contributions]  SN preformed a numerical calculations with some help from JT.  All authors developed the mathematical model,
   discussed the physical interpretation of the results and jointly
   wrote the manuscript.
 \item[Competing Interests] The authors declare that they have no
   competing financial interests.
 \item[Correspondence] Correspondence and requests for materials
   should be addressed to S.N \newline ~(email:
   snaoz@northwestern.edu).
\end{addendum}

 \newpage 

\vskip 0.2in \hrule \vskip 0.3in 
\centerline{\bf \large Supplementary Information}
\vskip 0.2in \hrule \vskip 0.3in 

\noindent\textbf{Octupole-order Evolution Equations and Angular
  Momentum Conservation }

Our  derivation corrects an error in previous Hamiltonian
derivations of the secular evolution equations.

We consider a hierarchical triple system consisting of an inner binary
($m_1$ and $m_2$) and a third body ($m_3$) in a wider exterior orbit.
We describe the system using canonical variables, known as Delaunay's
elements, which provide a particularly convenient dynamical
description of our three-body system. The coordinates are chosen to be
the mean anomalies, $l_1$ and $l_2$, the longitudes of ascending
nodes, $h_1$ and $h_2$, and the arguments of periastron, $g_1$ and
$g_2$, where subscripts $1,2$ denote the inner and outer orbits,
respectively.  Their conjugate momenta are:
\begin{eqnarray}
L_1&=&\frac{m_1 m_2}{m_1+m_2}\sqrt{k^2(m_1+m_2)a_1} \ , \\ \nonumber
L_2&=&\frac{m_3(m_1+ m_2)}{m_1+m_2+m_3}\sqrt{k^2(m_1+m_2+m_3)a_2} \ , 
\end{eqnarray}
\begin{equation}
G_1=L_1\sqrt{1-e_1^2} \ , \quad G_2=L_2\sqrt{1-e_2^2} \ ,
\end{equation}
where $k^2$ is the gravitational constant, and
\begin{equation}
H_1=G_1\cos{i_1} \ , \quad H_2=G_2\cos{i_2} \ ,
\end{equation}
where $G_1$ and $G_2$ are the absolute values of the angular momentum
vectors (${\bf G}_1$ and ${\bf G}_2$), and $H_1$ and $H_2$ are the
z-components of these vectors.

\begin{figure}[hptb]
\begin{center}
\includegraphics[width=0.5\textwidth]{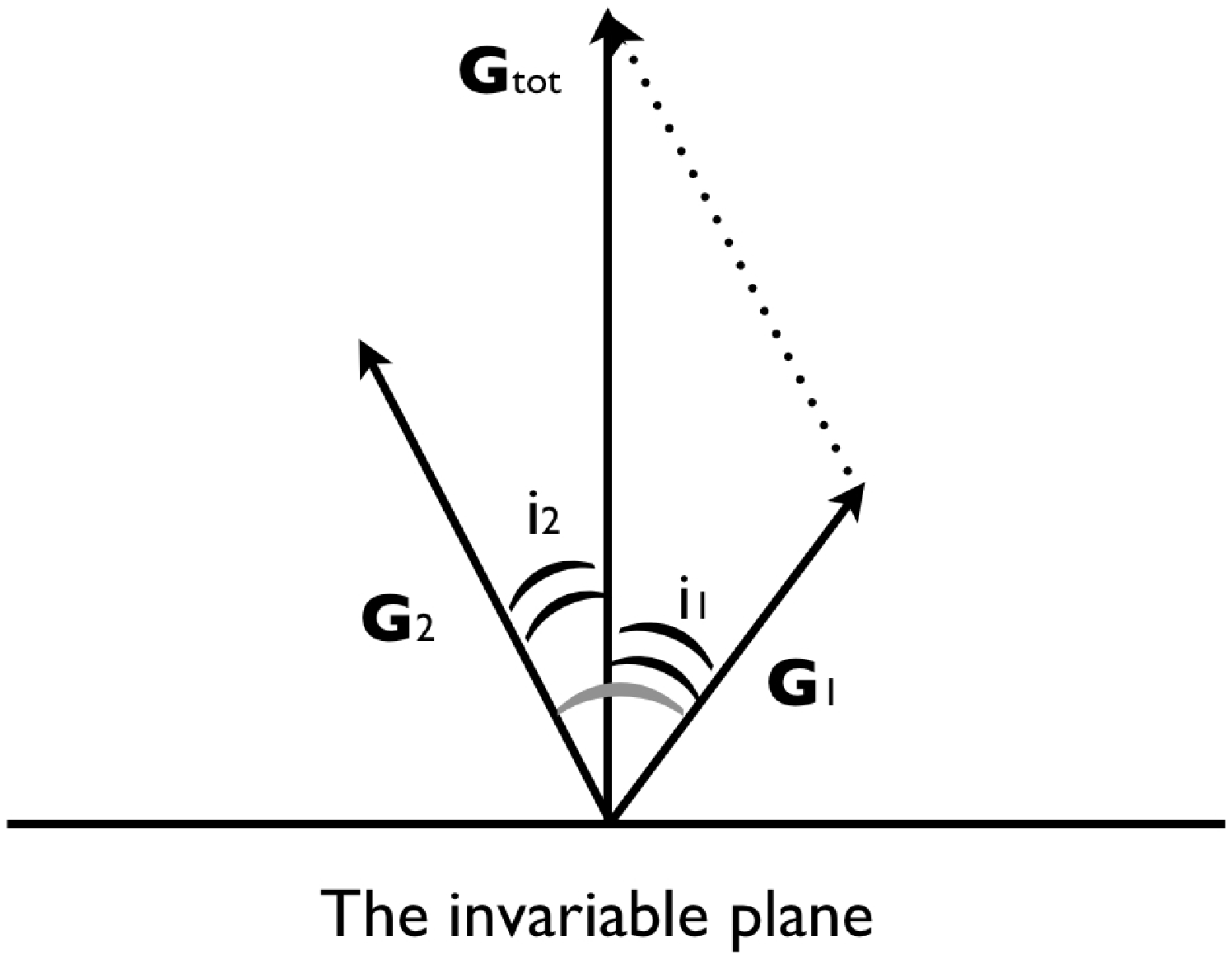}
\end{center}
\caption{\label{fig:S_angular} \textbf{\small The angular momenta of the
    bodies after the elimination of the nodes (see also Ref.~4).} Note
  that all three vectors are in the same plane. The mutual inclination
  $i = i_1 + i_2$ is the angle between ${\bf G}_1$ and ${\bf G}_2$.}
\end{figure}

We choose to work in a coordinate system where the total initial
angular momentum of the system lies along the $z$ axis.  The
transformation to this coordinate system is known as the elimination
of the nodes$^{30,17}$; the $x$-$y$ plane in this coordinate system is
known as the invariable plane.  Figure \ref{fig:S_angular} shows the
resulting configuration of the orbits.  We obtain simple relations
between $H_1$, $H_2$, $G_1$ and $G_2$, using ${\bf G}_{\tot}={\bf
  G}_1+{\bf G}_2$:
\begin{eqnarray}
\cos{i}&=&\frac{G_{\tot}^2-G_1^2-G_2^2}{2G_1G_2} \ , \label{eq:cosi} \\ 
H_1&=&\frac{G_{\tot}^2+G_1^2-G_2^2}{2G_{\tot}} \ , \label{eq:H1} \\ 
H_2&=&\frac{G_{\tot}^2+G_2^2-G_1^2}{2G_{\tot}} \ ,  \label{eq:H2}
\end{eqnarray}
where the relation for $H_1$ comes from setting ${\bf G}_2={\bf
  G}_{\tot} -{\bf G}_1$ (and similarly for $H_2$).  Because total
angular momentum is conserved by the evolution of the system, we must
have ${\bf G}_1(t) + {\bf G}_2(t) = {\bf G}_{\tot} = G_{\tot}
\hat{z}$, implying that 
\begin{equation}
h_1(t) = h_2(t) -\pi  \label{eq:Dh}.
\end{equation}

\vskip 0.3in

The Hamiltonian for the three-body system can be transformed into the form
\begin{equation}
  \Ham = \Ham^K_1(L_1) + \Ham^K_2(L_2) + \Ham_{12},
\end{equation}
where $\Ham^K_1$ and $\Ham^K_2$ represent the Keplerian interaction
between bodies 1 and 2 and the central body, and $\Ham_{12}$
represents the interaction between body 1 and body 2.  The Kepler
Hamiltonians depend only on the momenta $L_1$ and $L_2$, while the
interaction Hamiltonian, $\Ham_{12}$, depends on all the coordinates
and momenta.  Due to the rotational symmetry of the problem,
$\Ham_{12}$ depends on $h_1$ and $h_2$ only through the combination
$h_1 - h_2$.  Because we are interested in secular effects, we average
the Hamiltonian over the coordinates (angles) $l_1$ and $l_2$,
obtaining the secular Hamiltonian
\begin{equation}
  \bar{\Ham} = \Ham^K_1(L_1) + \Ham^K_2(L_2) + \bar{\Ham}_{12},
\end{equation}
where 
\begin{equation}
  \bar{\Ham}_{12} = \frac{1}{4\pi^2} \int_0^{2\pi} dl_1\, \int_0^{2\pi} dl_2\, \Ham_{12}.
\end{equation}

For simplicity we first focus on the quadrupole approximation, where
the error is more easily shown; it is then straightforward to see its
effects at all orders in the hierarchical triple system's secular
dynamics expansion.  The quadrupole Hamiltonian results from expanding
$\bar{\Ham}_{12}$ to second order%
\symbolfootnote[2]{The first order term in $a_1/a_2$ averages to zero, so the
  quadrupole term is the first term to contribute to
  $\bar{\Ham}_{12}$} %
in $a_1/a_2$:
\begin{equation}
  \bar{\Ham}_{12} = \bar{\Ham}^{(2)}_{12} + {\mathcal O}\left( \frac{a_1}{a_2} \right)^3.
\end{equation}
The resulting quadrupole-order Hamiltonian, $\bar{\Ham}^{(2)}_{12}$,
depends only on the coordinates $g_1$, $h_1$, and $h_2$, with the
latter two appearing only in the combination $h_1 - h_2$:
\begin{equation}
  \bar{\Ham}^{(2)}_{12} = \bar{\Ham}^{(2)}_{12}(g_1, h_1 - h_2).
\end{equation}
Previous calculations$^{17}$$^,$$^{20}$ eliminated $h_1$
and $h_2$ from the Hamiltonian using eq.~(\ref{eq:Dh}), obtaining a
quadrupole Hamiltonian that depends only on $g_1$.  But, this is
incorrect!  Such a Hamiltonian would imply that all quantities in
eq.~(\ref{eq:H1}) are constant {\em except} $G_1$, i.e.\ that
eq.~(\ref{eq:H1}) is incorrect.  Thus the previously used formalism
did not conserve angular momentum.  The initial Hamiltonian is
spherically symmetric, and therefore {\em does} conserve angular
momentum; the correct quadrupole Hamiltonian does as well.  Because
the correct quadrupole Hamiltonian depends on $h_1$ and $h_2$ through
the combination $h_1 - h_2$, we have
\begin{equation}
  \dot{H}_1 = - \dot{H}_2,
\end{equation}
or
\begin{equation}
  H_1 + H_2 = G_{\tot} = {\rm const}.
\end{equation}

The mathematical error affects all orders in secular perturbations.
The independence of the secular quadrupole Hamiltonian on $h_{1,2}$
was the source$^{17}$ of the famous relation  $\cos
i_{1,2}\sqrt{1-e_{1,2}^2}={\rm const}$. In the correct derivation,
this relation does not always hold.  However, in a certain limit, it
does.  From eq.~(\ref{eq:H1}), we see that
\begin{equation}
  \dot{H}_1 = \frac{G_1}{G_{\tot}} \dot{G}_1 - \frac{G_2}{G_{\tot}} \dot{G}_2.
\end{equation} 
When $G_2 \sim G_{\tot} \gg G_1$, we have
\begin{equation}
  \dot{H}_1 \approx -\frac{G_2}{G_{\tot}} \dot{G}_2.
\end{equation}
At the quadrupole level $\bar{H}^{(2)}_{12}$ is independent of $g_2$,
so $\dot{G}_2 = 0$, implying
\begin{equation}
  \dot{H}_1 \approx 0,
\end{equation}
when $G_2 \sim G_{\tot} \gg G_1$.  This is precisely the limit
considered in previous works$^{2,3,17,18,19,31}$, so their
conclusion that $H_{1,2} = \cos i_{1,2}\sqrt{1-e_{1,2}^2}={\rm const}$ is
correct (though not for the reason they claim), but the limit where
$G_2 \gg G_1$ is not sufficient for our work.

In some later studies, the assumption that $H_1={\rm const}$ was built
into the calculations of secular evolution for various astrophysical
systems$^9$$^,$$^{22-24}$,
even when the condition $G_2 \gg G_1$ was not satisfied. Moreover many
previous studies simply set $i_2=0$, which is repeating the same
error. In fact, given the mutual inclination $i$, the inner and outer
inclinations $i_1$ and $i_2$ are set by the conservation of total
angular momentum:
\begin{eqnarray}
\cos i_1&=&\frac{G_{\tot}^2+G_1^2-G_2^2}{2G_{\tot}G_1} \ , \label{eq:i1} \\ 
\cos i_2&=&\frac{G_{\tot}^2+G_2^2-G_1^2}{2G_{\tot}G_2} \ .  \label{eq:i2} 
\end{eqnarray}

\noindent\textbf{ Tidal Friction } 

We adopt the tidal evolution equations of Ref.~16, which are based on
the equilibrium tide model of Ref.~32. The complete equations
can be found in Ref.~2, eqs~A1--A5. Following their approach (see
their eq.~A10) we set the tidal quality factors $Q_{1,2}\propto P_{\rm
  in}$ [see also Ref.~33].  This means that the viscous
times of the star and planet remain constant; the representative
values we adopt here are $5$~yr for the star and $1.5$~yr for the
planet, which correspond to $Q_\star =5.5 \times 10^6$ and
$Q_J=5.8\times 10^6$, respectively, for a 1-day period.

\noindent\textbf{Comparison to Observations}

The observable parameter from the Rossiter--McLaughlin effect is the
{\em projected\/} angle between the star's spin and the orbital angular momentum
(the projected obliquity)$^{14}$.  Here instead we focus on the
true angle between the orbital angular momentum of the inner planet and the total angular momentum. 
Projection effects can cause these 
two quantities to differ in magnitude, or even sign.  

Moreover, several mechanisms have been proposed in the literature that could,
under certain assumptions, directly
affect the spin axis of the star. These mechanisms can re-align
the stellar spin axis through tidal interactions with either a
slowly spinning star$^{29}$ or with the outer convective layer
of a sufficiently cold star$^{10}$. Additionally,
a magnetic interaction between the star and the
protoplanetary disk could also lead 
to misalignment between the stellar spin and the disk$^6$. 

These effects can potentially complicate the interpretation of any specific
observation.  Nevertheless, if hot Jupiters are produced by the simple
mechanism described here, many of their orbits should indeed be observed with 
large projected obliquities.
%


\begin{thebibliography}{99}
\expandafter\ifx\csname url\endcsname\relax
  \def\url#1{\texttt{#1}}\fi
\expandafter\ifx\csname urlprefix\endcsname\relax\def\urlprefix{URL }\fi
\providecommand{\bibinfo}[2]{#2}
\providecommand{\eprint}[2][]{\url{#2}}

\vskip 30cm

\bibitem{Tri+10} Triaud, A. H. M. J., {\it et~al.}  Spin-orbit angle
  measurements for six southern transiting planets. New insights into
  the dynamical origins of hot Jupiters. {\it Astron. Astrophys.} {\bf 524}, A25 (2010).

\bibitem{Dan} Fabrycky, D. \& Tremaine, S.  Shrinking binary and
  planetary orbits by Kozai cycles with tidal friction.  {\it
    Astrophys. J.} {\bf 669}, 1298--1315 (2007).
  
\bibitem{Wu+07} Wu, Y., Murray, N.~W. \& Ramsahai, J.~M.  Hot Jupiters
  in binary star systems.  {\it Astrophys. J.} {\bf 670}, 820--825
  (2007).

\bibitem{Ford+00} Ford, E.~B., Kozinsky, B. \& Rasio, F.~A.  Secular
evolution of hierarchical triple star systems.  {\it Astrophys. J.}
{\bf 535}, 385--401 (2000).

\bibitem{Sourav+08} Chatterjee, S., Matsumura, S., Ford, E.~B. \&
  Rasio ,F.~A.  Dynamical outcomes of planet-planet scattering. {\it
    Astrophys. J.} {\bf 686}, 580--602 (2008).

\bibitem{Lai+10} Lai, D., Foucart, F. \& Lin, D.~N.~C.  Evolution of
  spin direction of accreting magnetic protostars and spin-orbit
  misalignment in exoplanetary systems.  {\it
    Mon. Not. R. Astron. Soc. (submitted);}  {\bf
    preprint at  http://arxiv.org/ abs/1008.3148}, (2011).

\bibitem{Nag+08} Nagasawa, M., Ida, S. \& Bessho, T. Formation of hot
  planets by a combination of planet scattering, tidal
  circularization, and the Kozai mechanism.  {\it Astrophys. J.} {\bf
    678}, 498--508 (2008).

\bibitem{Schlaufman} Schlaufman, K.~C.  Evidence of possible
  spin-orbit misalignment along the line of sight in transiting
  exoplanet systems.  {\it Astrophys. J.} {\bf 719}, 602--611 (2010).
  
\bibitem{Takeda} Takeda, G., Kita, R. \& Rasio, F.~A.  Planetary
  systems in binaries. I. Dynamical classification.  {\it
    Astrophys. J.} {\bf 683}, 1063--1075 (2008).

\bibitem{Winn+10b} Winn, J.~N., Fabrycky, D., Albrecht, S. \& Johnson,
  J.~A. Hot stars with hot Jupiters have high obliquities. {\it Astrophys. J.}
 {\bf 718}, L145--L149 (2010).

\bibitem{WL} Wu, Y. \& Lithwick, Y.  Secular chaos and the production
  of hot Jupiters.  {\it preprint at http://arxiv.org/abs/1012.3475}
  (2010).
  
\bibitem{Lin+86} Lin, D.~N.~C. \& Papaloizou, J.  On the tidal
  interaction between protoplanets and the protoplanetary disk. III -
  Orbital migration of protoplanets.  {\it Astrophys. J.}  {\bf 309},
  846--857 (1986).
  
\bibitem{Mass+03} Masset, F.~S. \& Papaloizou, J.  Runaway migration
  and the formation of hot Jupiters.  {\it Astrophys. J.} {\bf 588},
  494--508 (2003).
  
\bibitem{GW07} Gaudi, B.~S. \& Winn, J.~N. Prospects for the
  characterization and confirmation of transiting exoplanets via the
  Rossiter-McLaughlin effect.  {\it Astrophys. J.} {\bf 655}, 550--563
  (2007).

\bibitem{Hol+97} Holman, M., Touma, J. \& Tremaine, S.  Chaotic
  variations in the eccentricity of the planet orbiting 16 Cygni B.
  {\it Nature} {\bf 386}, 254--256 (1997).

\bibitem{1998KEM} Eggleton, P.~P., Kiseleva, L.~G. \& Hut P.  The
  equilibrium tide model for tidal friction.  {\it Astrophys, J.} {\bf
    499}, 853--870 (1998).

\bibitem{Kozai} Kozai, Y.  Secular perturbations of asteroids with
  high inclination and eccentricity.  {\it Astron, J.} {\bf 67},
  591--598 (1962).
  
\bibitem{Lidov} Lidov, M.~L.  The evolution of orbits of artificial
  satellites of planets under the action of gravitational
  perturbations of external bodies.  {\it Planet. Space Sci.} {\bf 9},
  719--759 (1962).
  
\bibitem{Mazeh+79} Mazeh, T. \& Shaham, J.  The orbital evolution of
  close triple systems - The binary eccentricity.  {\it
    Astron. Astrophys.} {\bf 77}, 145--151 (1979).

\bibitem{Harr69} Harrington, R.~S.  The stellar three-body problem.
  {\it Celestial Mechanics} {\bf 1}, 200--209 (1969).

\bibitem{Mazeh+99} Krymolowski, Y. \& Mazeh, T.  Studies of multiple
  stellar systems - II. Second-order averaged Hamiltonian to follow
  long-term orbital modulations of hierarchical triple systems.  {\it
    Mon. Not. R. Astron. Soc.} {\bf 304}, 720--732 (1999).

\bibitem{1998EKH} Kiseleva, L.~G., Eggleton, P.~P. \& Mikkola, S.
  Tidal friction in triple stars.  {\it Mon. Not. R. Astron. Soc.}
  {\bf 300}, 292--302 (1998).

\bibitem{Zdz+07} Zdziarski, A.~A., Wen, L. \& Gierli{\'n}ski, M.  The
  superorbital variability and triple nature of the X-ray source 4U
  1820-303.  {\it Mon. Not. R. Astron. Soc.} {\bf 377}, 1006--1016
  (2007).

\bibitem{Mik+98} Mikkola, S. \& Tanikawa, K.  Does Kozai resonance
  drive CH Cygni?  {\it Astron. J.} {\bf 116}, 444--450 (1998).

\bibitem{Ford+06} Ford, E.~B. \& Rasio, F.~A.  On the relation between
  hot Jupiters and the Roche limit.  {\it Astron. J.} {\bf 638},
  L45--L48 (2006).

\bibitem{Kal+08} Kalas, P., {\it et~al.}  Optical images of an
  exosolar planet 25 light-years from Earth.  {\it Science} {\bf 322},
  1345--1348 (2008).

\bibitem{Mar+08} Marois, C., {\it et~al.}  Direct imaging of multiple
  planets orbiting the star HR 8799.  {\it Science} {\bf 322},
  1348--1352 (2008).

\bibitem{Pol+96} Pollack, J.~B., {\it et. al.}  Formation of the giant
  planets by concurrent accretion of solids and gas.  {\it Icarus}
  {\bf 124}, 62--85 (1996).
  
  
\bibitem{Soko} Matsumura, S., Peale, S.~J. \& Rasio, F.~A.  Formation
  and Evolution of Close-in Planets.  {\it Astrophys. J.} {\bf 725},
  1995--2016 (2010).



\bibitem{JM66} Jefferys, W.~H. \& Moser, J.  Quasi-periodic
  solutions for the three-body problem.  {\it Astron. J.} {\bf 71},
  568--578 (1966).
  
\bibitem{PN} Perets, H~B. \& Naoz, N. Tidal friction, and the
  dynamical evolution of binary minor planets.  {\it Astron. J.} {\bf
    699}, L17--L21 (2009).
  
\bibitem{Hut} Hut, P.  Tidal evolution in close binary systems.
  {\it Astron. Astrophys} {\bf 99}, 126--140 (1981).
  
\bibitem{Hansen} Hansen, P.  Calibration of equilibrium tide
  theory for extrasolar planet systems.  {\it Astrophys. J.} {\bf 723},
  285--299 (2010).
  

\end{thebibliography}
\end{document}